\documentclass[lettersize,journal]{IEEEtran}
\usepackage{amsmath,amsfonts}
\usepackage{algorithmic}
\usepackage{algorithm}
\usepackage{array}
\usepackage[caption=false,font=normalsize,labelfont=sf,textfont=sf]{subfig}
\usepackage{textcomp}
\usepackage{stfloats}
\usepackage{url}
\usepackage{verbatim}
\usepackage{graphicx}
\usepackage{cite}
\usepackage{url}
\usepackage{xurl}
\usepackage{multirow}
\usepackage{enumitem}
\begin{document}

\title{MIDS: Detecting Stealthy Masquerade and Tampering\\ Attacks on CAN Bus via Bidirectional Mamba}

\author{Qiqi~Liu, 
        Runhan~Song, 
        Lei~Cui,
        Heng~Zhang, 
        Yuyan~Sun, 
        and~Limin~Sun
\thanks{Qiqi Liu and Heng Zhang are with the Institute of Information Engineering, Chinese Academy of Sciences, Beijing 100093, China, and also with the School of Cyber Security, University of Chinese Academy of Sciences, Beijing 100049, China (e-mail: vrmeies@gmail.com; zhangheng@iie.ac.cn).}

\thanks{Yuyan Sun and Limin Sun are with the Institute of Information Engineering, Chinese Academy of Sciences, Beijing 100093, China (e-mail: sunyuyan@iie.ac.cn; sunlimin@iie.ac.cn).}

\thanks{Runhan Song and Lei Cui are with the Zhongguancun Laboratory, Beijing, China (e-mail: songrh2025@mail.zgclab.edu.cn; cuilei@mail.zgclab.edu.cn). }

\thanks{(Corresponding author: Yuyan Sun.)}
}

\maketitle

\begin{abstract}
The Controller Area Network (CAN) protocol is the primary 
communication standard for Electronic Control Units (ECUs) in 
modern vehicles, but its lack of encryption and authentication 
exposes it to a range of security threats. Existing intrusion 
detection systems are largely tuned to fabrication-style attacks 
(DoS, fuzzing, ID spoofing realised by frame injection), in which 
detection signals such as per-ID inter-arrival statistics are 
readily available. We instead address the harder \emph{masquerade} 
setting~\cite{b37}, in which an internal adversary substitutes a 
legitimate frame in-situ at its original transmission slot, 
preserving traffic periodicity and rendering traffic-statistic 
defences ineffective. We propose the Mamba Intrusion Detection 
System (MIDS), an innovative dual-stream framework that processes 
CAN identifiers and payloads in parallel and reconstructs their 
joint temporal semantics through bidirectional selective state-space 
modelling. To evaluate MIDS, we collected over 100 million CAN 
frames from a physical Tesla Model 3 across three driving regimes 
and synthesised 54 masquerade attack variants spanning ID-only, 
data-only, and combined modifications. MIDS attains an F1 of 96.94\% 
on this dataset, exceeding the strongest reproducible baseline by 
more than 8 percentage points, while sustaining a 1.147~ms 
single-window inference latency---ample headroom for real-time 
onboard deployment. To verify generalisation, we further evaluate
MIDS on four public benchmarks (ROAD, CrySyS, OTIDS, CT\&T)
covering both masquerade and injection scenarios; MIDS
attains F1 from 93.70\% to 99.61\%, outperforming the
strongest of eight reproduced baselines by up to 13.94
percentage points under a unified 5-fold protocol.
\end{abstract}


\section{Introduction}

\begin{figure*}[ht!] 
\centerline{\includegraphics[width=1.00\textwidth]{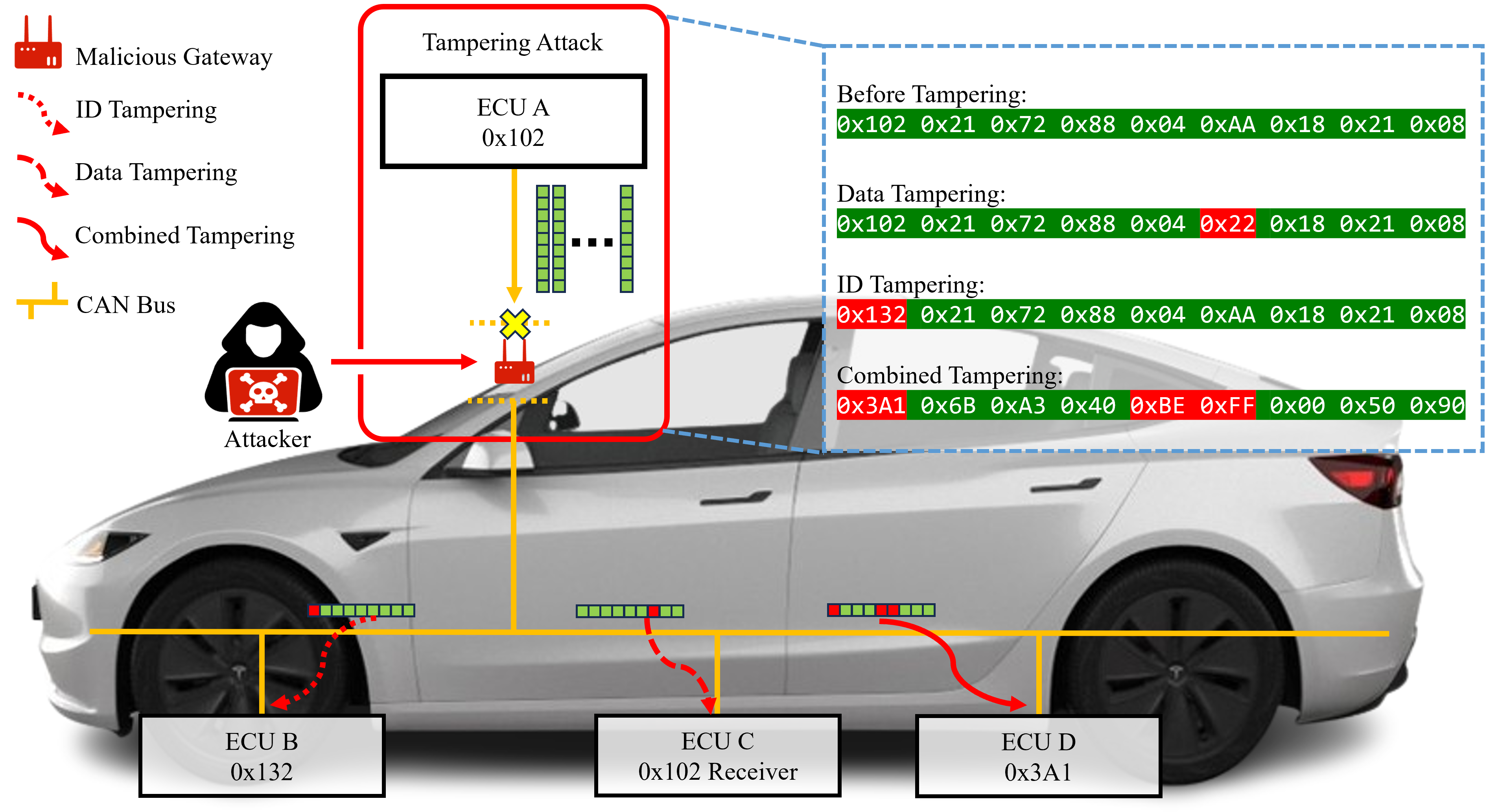}}
\caption{Overview of tampering attacks threat model. Attackers can exploit vulnerabilities in a weak gateway to initiate the entire tampering attack process. The compromised ECU, originally responsible for sending CAN frames with identifier ID 0x102, allows attackers to choose from three types of attacks: (1) Data tampering, (2) ID tampering, and (3) Combined tampering. Each type leads to severe consequences but impacts different targets. In (1), the attacker directly tampers with the data field of the CAN frame being sent, which affects the receiver of ID 0x102 (shown as the ECU C). This may result in scenarios such as a vehicle continuing to drive with its door open. In (2) and (3), the attacker tampers with the frame data to impersonate another ECU (e.g., sending IDs 0x132 or 0x3A1, shown as ECU B and D), thereby sending malicious data that could cause abnormal steering angles, compromising the entire vehicle system's safety. Unlike injection attacks, tampering attacks generally do not significantly impact traffic distribution because the overall traffic flow in the system remains unchanged.} 
\label{threat} 
\end{figure*}

\subsection{Background and Challenge}
The Controller Area Network (CAN) is a serial communication protocol widely used in automotive and embedded systems. For example, CAN was commonly adopted as the standard communication protocol of Electronic Control Units (ECUs) in modern vehicles. However, the protocol's original design did not account for security, resulting in the absence of modern security mechanisms such as encryption, authentication, and integrity verification\cite{b2,b5}. Consequently, if an attacker exploits a vulnerability in a specific ECU, they could potentially compromise the entire CAN bus\cite{b1}, allowing the execution of high-risk attacks.

In recent years, significant advancements have been made in detecting data tampering attacks \cite{b51,b52} and injection attacks \cite{b9,b10,b18,b14,b16,b17} in the Controller Area Network (CAN). While these methods have demonstrated effectiveness, attackers can exploit a new class of masquerade attacks \cite{b37}, which pose a greater threat, provided that they have gained access to the network gateway. Following the naming convention and threat model established in \cite{b37}, we define this masquerade attack as a sophisticated form of tampering that alters the CAN message’s ID field, effectively impersonating a legitimate ECU to compromise the integrity of the system. When traffic passes through the gateway, it may be intercepted and modified by attackers without violating the bus's timing constraints \cite{b37}.

In this paper, we adopt a comprehensive threat model that encompasses masquerade (ID tampering), data tampering, and combined tampering attacks. As illustrated in Fig.~\ref{threat}, these attacks typically involve altering critical data (e.g., sensor readings or control commands) in order to maliciously manipulate device behavior (e.g., directing an ECU to send spoofed commands). Such manipulations can lead to severe and potentially catastrophic consequences.

While previous research has extensively covered injection attacks, stealthy masquerade and tampering attacks—which manipulate internal message fields—have received far less attention. In particular, the complex interplay between ID and data field tampering remains a significant yet underexplored challenge. Our work systematically addresses this gap by proposing a threat model that encompasses masquerade, data, and hybrid tampering scenarios. These attacks are inherently difficult to detect because they rely on in-situ modifications that maintain the original traffic periodicity and frequency. Consequently, traditional Intrusion Detection Systems (IDS) focusing on traffic patterns are often ineffective against these subtle semantic-level anomalies \cite{b52}.

Furthermore, these masquerade attacks are considered highly feasible. As highlighted in prior studies \cite{b61}, attackers can carry out such tampering by compromising the gateway adjacent to the sender and then intercept and modify CAN messages, thereby manipulating data on the CAN bus. In summary, it is crucial to develop advanced feature extraction techniques and deep learning models capable of capturing the subtle, often undetectable characteristics of masquerade attacks, including both ID and data modifications.

\subsection{Contribution}
To address the challenges posed by these stealthy threats, we propose the Mamba Intrusion Detection System (MIDS), a robust solution engineered to capture the subtle semantic-level anomalies inherent in various masquerade (ID-level) and tampering (data-level) scenarios. We evaluate MIDS using an extensive dataset collected from a physical Tesla Model 3 and several publicly available benchmarks, covering masquerade, data, and hybrid tampering attacks, alongside traditional injection attacks. To foster reproducibility and facilitate further research, the source code of MIDS and the collected dataset have been made publicly available\footnote{The source code is available at \url{https://github.com/vrmei/MIDS}, and the dataset can be accessed at \url{https://drive.google.com/drive/folders/13mutfkDNRI3iIHZkrwS2YLVDmgfDYeTV?usp=drive_link}.}. Our main contributions are summarized as follows:
\begin{itemize}
    \item Firstly, we introduce a comprehensive threat model, as shown in Fig.~\ref{threat}, which encompasses masquerade attacks alongside data and hybrid tampering scenarios. Compared to previous works, these combined attacks are covert and difficult to detect because they do not alter traffic patterns. Additionally, ID tampering attacks have a broader attack surface, making them more flexible and potentially more harmful. The new threat model provides a realistic simulation of attacker behavior, presenting significant challenges for Intrusion Detection Systems (IDS) in detecting such attacks.
    
    \item Secondly, we propose MIDS, an innovative framework that integrates the Mamba state-space model with Convolutional Neural Networks (CNNs). To our knowledge, this is the first application of Mamba to CAN-bus masquerade and tampering detection. By leveraging the linear complexity and superior long-sequence modeling of Mamba, MIDS effectively captures high-dimensional temporal characteristics and cross-field correlations with high computational efficiency, making it particularly suitable for the resource-constrained embedded environments of modern vehicles. Experimental results demonstrate that MIDS significantly outperforms existing methods across key performance metrics, including accuracy and F1 score. Under a unified 5-fold protocol on five datasets, MIDS exceeds the strongest of eight reproduced baselines by up to 13.94 percentage points.
    
    \item Thirdly, to evaluate the effectiveness of MIDS in detecting tampering attacks, we collected a comprehensive dataset of real-world CAN data from a physical Tesla Model 3, covering a range of driving scenarios, including highway driving, urban driving, parking, and braking. This dataset fills a critical gap in existing benchmarks like CrySyS\cite{b33} and ROAD\cite{b34}, which lack stealthy masquerade and hybrid tampering attacks. The dataset contains masquerade, data tampering, and combined tampering attacks of varying intensities. Additionally, we incorporated four other public datasets\cite{b20,b28,b33,b34} to further assess the robustness of MIDS against traditional injection attacks. To foster future research and collaboration, both our dataset and MIDS are publicly released, providing a solid foundation for advancing vehicular cybersecurity.
    
\end{itemize}
\begin{table*}[htbp]
\caption{Summary of CAN datasets}
\label{tab2}
\begin{center}
\renewcommand{\arraystretch}{1.2} % Adjust row height
\small
\begin{tabular}{cccccc}
\hline
\textbf{Name} & \textbf{Years} & \textbf{Rows} & \textbf{Attack Type} & \textbf{Reality} & \textbf{Vehicle Model} \\
\hline
Simulated CAN\cite{b19} & 2016 & 200,000 & Injection & Simulated & N/A \\
HCRL CAN (OTIDS)\cite{b20} & 2017 & 4,613,909 & Injection & Real & KIA SOUL \\
HCRL Car-Hacking\cite{b21} & 2018 & 17,558,462 & Injection & Real & YF Sonata \\
AEGIS CAN\cite{b22} & 2019 & 3,462,015 & Benign only & Real & Unknown \\
Bus-Off\cite{b23} & 2019 & 189,083,068 & Injection & Simulated & Volvo V40 \\
TU CAN v2\cite{b24} & 2019 & 11,830,305 & Injection & Real & Opel and Renault \\
ML350 CAN\cite{b25} & 2019 & 730,519 & Injection & Real & ML350 \\
ReCAN\cite{b26} & 2020 & 38,000,000 & Benign only & Real & 5 Unknown Vehicles  \\
SynCAN\cite{b27} & 2020 & 42,958,391 & Injection & Simulated & Unknown \\
HCRL A\&D\cite{b29} & 2020 & 8,694,507 & Injection & Real & Avante CN7 \\
Truck CAN Dataset\cite{b30} & 2021 & 530,810,616 & Benign only & Real & Renault Euro VI \\
ROAD CAN Dataset\cite{b34} & 2021 & ~1.1M & Injection \& Masquerade & Real \& Simulated & Unknown(2010s) \\
DAGA\cite{b31} & 2022 & 200,000,000 & Injection & Real \& Simulated & N/A \\
Ventus\cite{b32} & 2023 & 539,657,925 & Injection & Simulated & N/A \\
CT\&T\cite{b28} & 2023 & 193,241,081 & Injection & Real \& Simulated & Multiple Chevrolet \\
CrySyS CAN\cite{b33} & 2023 & 138,362,148 & Injection \& Masquerade & Real \& Simulated & Unknown \\
Ours & 2024 & 108,053,935 & Masquerade & Real \& Simulated & Tesla Model 3 \\
\hline
\end{tabular}
\end{center}
\end{table*}

\section{Related Work}
\subsection{CAN Intrusion Detection Methodologies}
In recent studies, researchers have focused on the CAN protocol's fixed ID frequency and the periodic message transmission by ECUs. \cite{b7} proposed a frequency distribution-based intrusion detection method to monitor frequency variations, while \cite{b35,b42} introduced entropy-based anomaly detection to capture sequence uncertainties. Overall, most early works relied on statistical features \cite{b39, b40, b41, b47} and protocol specifications. Most notably, the seminal work by Cho and Shin \cite{b37} leveraged protocol-level characteristics and physical-layer clock skews to implement ECU fingerprinting. This study not only enhanced detection through protocol compliance but also established the standard naming conventions for in-vehicle attacks—particularly defining "masquerade attacks"—which provides the foundational terminology for analyzing stealthy threats that maintain original traffic periodicity.

Although these methods can identify anomalous behaviors to some extent, they struggle with complex, non-periodic attacks. Consequently, machine learning (ML) and deep learning (DL) techniques have gained prominence. ML-based systems, such as those using Random Forest and SVM \cite{b38}, offer greater flexibility than rule-based methods but often require manual feature engineering.

With the rapid development of DL, its capabilities in feature extraction and time-series processing have introduced new paradigms. For instance, \cite{b19} applied Deep Belief Networks (DBN) to classify messages, while \cite{b43} leveraged LSTM networks to predict temporal sequences. Generative approaches (GANs) \cite{b21} and federated learning \cite{b46, b47} have also been explored. Additionally, CNN-based architectures \cite{b48, b49, b50} have been widely adopted to capture spatial features within CAN data. However, as these models grow in complexity, balancing detection accuracy with real-time hardware constraints becomes a critical challenge. To address this, recent work has explored specialized 
architectures along several complementary axes. Khandelwal 
and Shreejith~\cite{b62} and Rangsikunpum 
\textit{et al.}~\cite{b63} target 
deployment cost, mapping CNN- and binarised-neural-network 
detectors onto low-cost FPGAs to achieve sub-millisecond 
inference at the expense of model expressiveness. Park 
\textit{et al.}~\cite{b64} reformulate detection 
as a graph-classification problem over message dependency 
graphs, capturing topological regularities of injection 
traffic. Ma \textit{et al.}~\cite{b65} pursue runtime 
efficiency through a lightweight GRU detector tailored for 
real-time onboard inference.

However, these works are evaluated exclusively against 
fabrication-style attacks, where per-ID inter-arrival 
statistics provide the primary detection signal. Under the 
masquerade model adopted in this paper 
(Section~\ref{sec:threat}), this signal vanishes by 
construction, motivating the dual-stream bidirectional 
design of MIDS.

For data tampering detection, \cite{b51} proposed a protocol embedding partial MAC values, while \cite{b52} employed MS-iForest to improve sensitivity to local anomalies. However, while these approaches provide effective strategies, they focus predominantly on isolated data field modifications and struggle to capture the complex, cross-field semantic dependencies inherent in sophisticated masquerade attacks. Specifically, protocol-based methods \cite{b51} introduce significant bus overhead in high-frequency scenarios, while statistical methods \cite{b52} often lack the temporal depth to distinguish stealthy tampering from normal signal fluctuations.

For completeness, we note that ML-based IDS is not the only
class of defence against tampering. Statistical detectors based
on Hamming distance~\cite{b66}, n-gram analysis~\cite{b31},
or CAN-signal entropy~\cite{b67} offer lightweight, training-free
alternatives, while cryptographic protections---most notably the
AUTOSAR SecOC standard~\cite{b68} and recent zero-overhead
schemes such as ZBCAN~\cite{b69}---attach MAC-based
authentication to each frame to prevent tampering at the
protocol level. These approaches are complementary rather than
substitutive: cryptographic defences require coordinated OEM
deployment and key-management infrastructure that has not yet
reached mass adoption, while statistical detectors trade
expressive power for simplicity. MIDS targets the deployable
middle ground---a learned IDS that runs on already-deployed
gateway hardware without requiring protocol upgrades---and our
benchmark comparison (Section~\ref{sec:exp_public}) is therefore
scoped to other learned IDS baselines under the same input
interface. Within this learned-IDS class, the choice of Mamba
over alternative sequence backbones (LSTM, Transformer) is
empirically motivated: Mamba's selective state-space mechanism
preserves the linear-time complexity of RNNs while capturing the
long-range dependencies traditionally accessible only through
quadratic-cost attention, a trade-off we quantify in
Sections~\ref{sec:sota_comp}--\ref{sec:efficiency}.

\subsection{In-Vehicle Security Benchmarks}

We surveyed existing CAN-bus attack datasets in
Table~\ref{tab2}. Two limitations are common across the prior
landscape. First, the vast majority target injection attacks
(DoS, fuzzing, replay, ID spoofing) whose detection relies on
traffic-statistic shortcuts; only ROAD and CrySyS include
tampering scenarios, and even those focus on data-field anomalies
rather than the ID-level masquerade we study. Second, most
publicly available datasets are collected from older vehicle
platforms (KIA Soul, YF Sonata, Opel/Renault, etc.) and do not
reflect the centralized E/E architectures of modern electric
vehicles.

Our Tesla Model~3 dataset directly addresses both gaps: it
combines high-fidelity physical traces from a contemporary EV
with 54 professionally crafted masquerade and tampering attack
variants, providing the first publicly released benchmark on
which IDS robustness against stealthy semantic-level anomalies
can be rigorously evaluated.

\section{Preliminaries}
\subsection{CAN Frame Structure}
\begin{figure}[h]
\centerline{\includegraphics[width=\columnwidth]{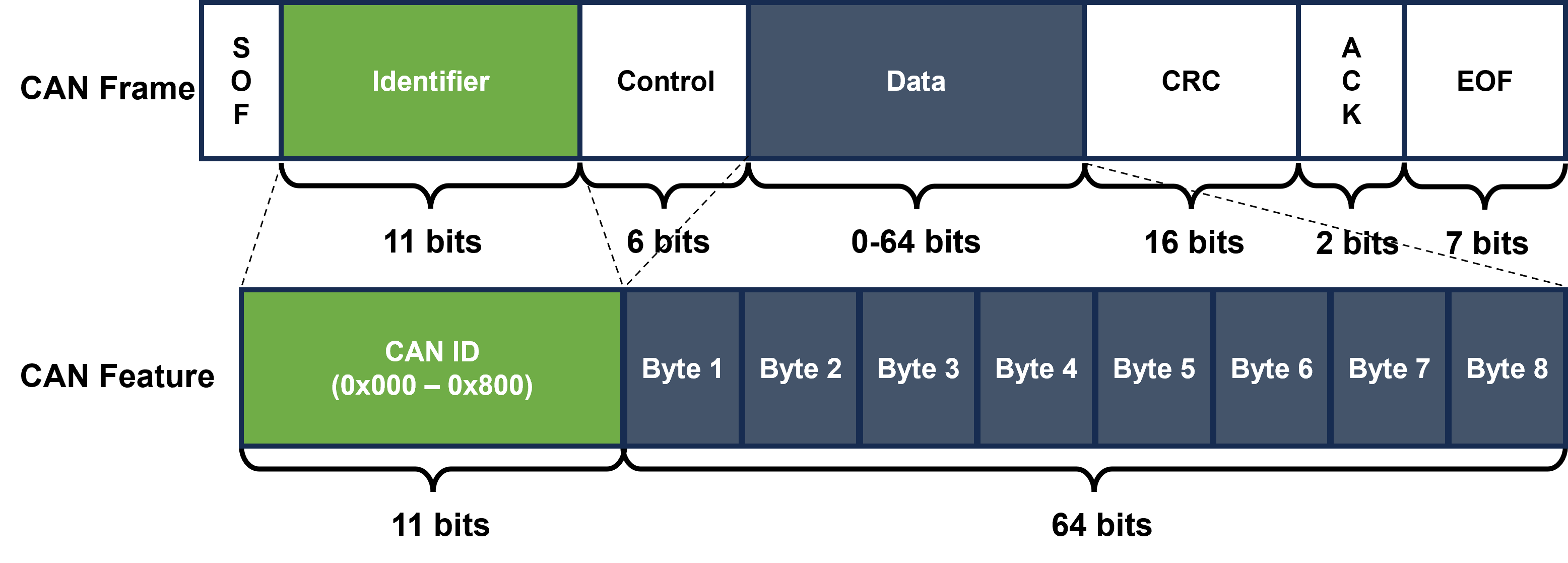}}
\caption{CAN frame structure}
\label{CanFrame}
\end{figure}

As illustrated in Fig.~\ref{CanFrame}, a standard CAN 2.0B frame consists of several functional fields that ensure reliable communication across the vehicular network. To understand the mechanism of stealthy attacks, it is essential to distinguish between the semantic payloads and the protocol control fields:
\begin{itemize}[leftmargin=*, nosep]
\item \textbf{Start of Frame (SOF):} A single dominant bit marking the beginning of a transmission.
\item \textbf{Identifier (ID) Field:} Defines the message priority and its functional category. In masquerade attacks, this field is manipulated to impersonate a legitimate ECU while maintaining the original transmission frequency.
\item \textbf{Control Field:} Contains the Data Length Code (DLC), specifying the number of bytes in the data field.
\item \textbf{Data Field:} The core payload containing sensor values or control commands (up to 8 bytes). Tampering attacks primarily target this field to inject malicious semantic information.
\item \textbf{Cyclic Redundancy Check (CRC):} A 15-bit parity bit used for detecting transmission errors. Importantly, in modern E/E architectures, the CRC is automatically generated and verified by the CAN controller hardware at the Data Link Layer. Thus, an internal attacker who has compromised an ECU’s software stack can broadcast malicious frames with valid CRCs, bypassing traditional integrity checks.
\item \textbf{Acknowledgment (ACK) Bit:} Used by receivers to confirm the successful reception of a valid frame.
\item \textbf{End of Frame (EOF):} A 7-bit sequence signaling the completion of the frame.
\end{itemize}

\subsection{Selective State Space Models (Mamba)}

State Space Models (SSMs) \cite{b53} are defined by a continuous system that maps an input signal $u(t) \in \mathbb{R}$ to an output $y(t) \in \mathbb{R}$ through a latent state $h(t) \in \mathbb{R}^N$:
\begin{equation}
\dot{h}(t) = \mathbf{A}h(t) + \mathbf{B}u(t), \quad y(t) = \mathbf{C}h(t)
\label{eq:ssm_continuous}
\end{equation}
where $\mathbf{A}, \mathbf{B}, \mathbf{C}$  are evolution parameters. To process discrete sequence data $x = \{x_0, x_1, \dots\}$, Eq. \eqref{eq:ssm_continuous} is discretized using a step size $\Delta$. Typically, the zero-order hold (ZOH) rule is applied to transform $(\mathbf{A}, \mathbf{B})$ into discrete parameters $(\mathbf{\bar{A}}, \mathbf{\bar{B}})$:
\begin{equation}
\mathbf{\bar{A}} = \exp(\Delta \mathbf{A}), \quad \mathbf{\bar{B}} = (\Delta \mathbf{A})^{-1}(\exp(\Delta \mathbf{A}) - \mathbf{I}) \cdot \Delta \mathbf{B}
\label{eq:discretization}
\end{equation}
While structured SSMs allow for efficient $O(N)$ inference and $O(N \log N)$ training via convolution, they suffer from parameter rigidity, failing to perform context-aware information filtering. Mamba \cite{b54} addresses this by introducing a Selective Mechanism (S6), where the parameters $(\mathbf{B, C, \Delta})$ are formulated as data-dependent functions:
\begin{equation}
\begin{aligned}
\mathbf{B} &= \text{Linear}_B(x), \\
\quad \mathbf{C} &= \text{Linear}_C(x), \\
\quad \Delta &= \text{Softplus}(\text{Parameter} + \text{Linear}_\Delta(x))
\label{eq:selection}
\end{aligned}
\end{equation}
By making these parameters functions of the input $x_t$, Mamba enables the model to selectively propagate or "forget" information based on the current context. This selection mechanism maintains the linear computational complexity of RNNs while achieving the modeling power of Transformers, making it uniquely suited for capturing the long-range temporal dependencies in CAN message sequences.

\begin{figure*}[htbp]
    \centering
    \includegraphics[scale=0.6]{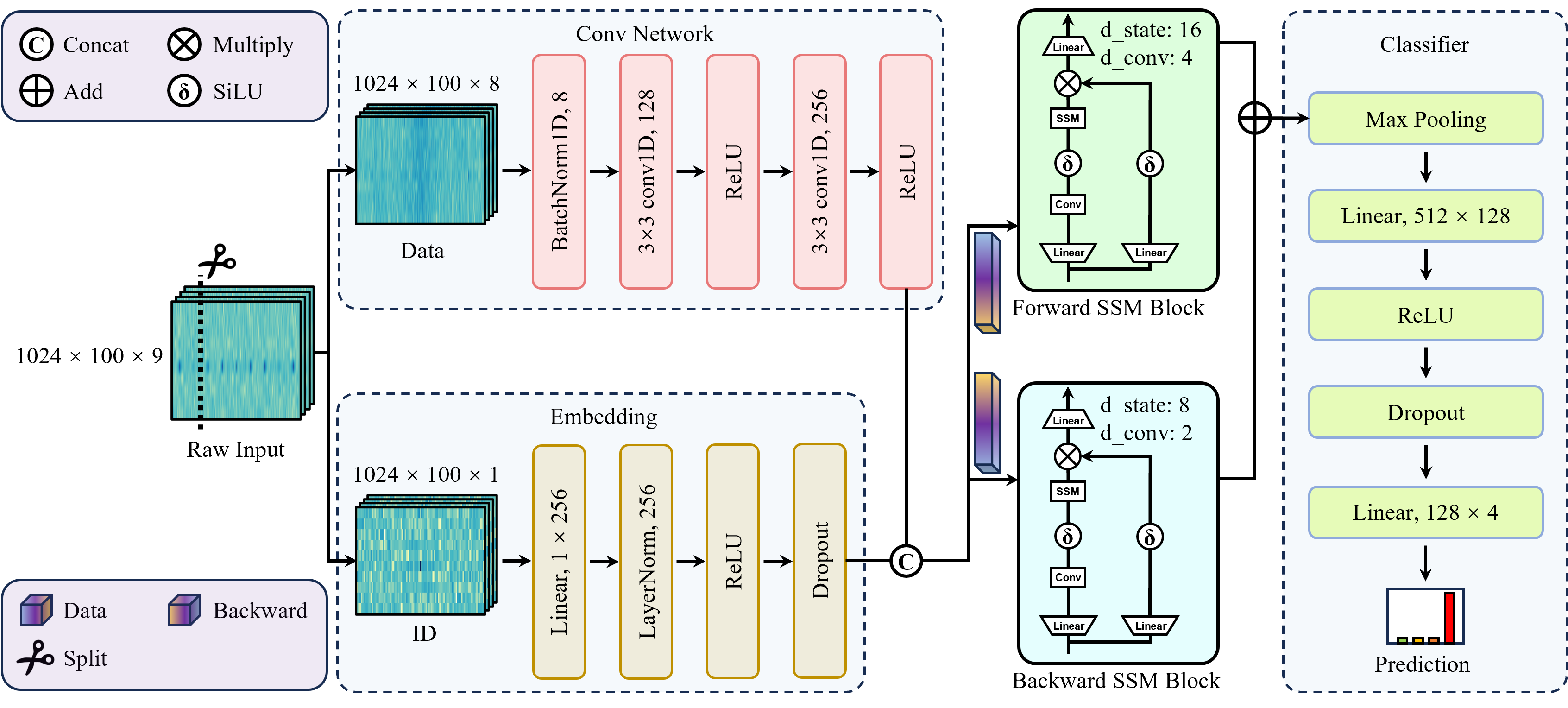}
    \caption{Model architecture of MIDS. The Forward and Backward SSM
    blocks adopt an asymmetric configuration ($d_{\text{state}}=16,
    d_{\text{conv}}=4$ for the forward branch; $d_{\text{state}}=8,
    d_{\text{conv}}=2$ for the backward branch); see
    Section~\ref{sec:bimamba} for the rationale.}
    \label{model}
\end{figure*}

\section{Threat Model and Attack Taxonomy}
\label{sec:threat}

This section formalizes the adversarial setting under which 
MIDS is designed and evaluated. We adopt the masquerade-attacker 
taxonomy introduced by Cho and Shin~\cite{b37}, 
which distinguishes \emph{fabrication} attacks (in which the 
adversary injects additional frames and thereby disturbs 
per-ID periodicity) from \emph{masquerade} attacks (in which 
the adversary suppresses a legitimate sender and substitutes 
a crafted frame at its original timeslot, preserving 
periodicity by construction). All three attack vectors 
considered in this paper---ID modification, data-field 
modification, and their combination---are instances of the 
masquerade pattern: the malicious frame replaces a legitimate 
one rather than being appended to the bus, and consequently 
none of them produces the inter-arrival-time disturbance that 
fabrication-oriented IDS rely on. The overall scenario is 
illustrated in Fig.~\ref{threat}.

\subsection{Attacker Capabilities and Assumptions}
Unlike traditional external attackers, we assume a high-privilege adversary who has gained access to the vehicle's internal network via a compromised Telematics Control Unit (TCU) or a malicious diagnostic tool. Specifically, the attacker is assumed to have:
\begin{itemize}[leftmargin=*, nosep]
\item \textbf{Software-level Control:} The attacker has compromised the central gateway or a high-priority ECU's software stack. As shown in Fig. 1, this allows the attacker to intercept legitimate traffic from ECU A before it reaches the broader CAN bus.
\item \textbf{In-situ Modification:} Instead of injecting redundant frames (which would increase bus load and be easily detected by frequency-based IDS), the attacker performs \textit{in-situ} substitution. The malicious gateway suppresses the original frame and immediately broadcasts a modified version (ID, Data, or both) at the exact same timestamp.
\item \textbf{Hardware-assisted CRC:} A critical aspect of our model is the bypass of integrity checks. Since the modification occurs at the application/software layer of the compromised node, the on-chip CAN controller hardware automatically calculates and appends a valid CRC for the tampered frame. Consequently, receiving nodes (ECUs B, C, and D) accept the malicious messages as protocol-compliant.
\end{itemize}

We emphasise that the attacker capabilities above describe the
\emph{threat model} underlying our evaluation; the corresponding
attack traces are synthesised offline by replacing or modifying
frames in benign captures collected from a physical Tesla Model~3,
rather than by actively manipulating a live bus. This choice
provides reproducible, byte-exact ground-truth labels for the
$54$ attack variants in Section~\ref{sec:dataset_design} and
sidesteps the orthogonal engineering problem of real-time
dominant-bit overwriting, which prior work has shown to be
feasible only with dedicated hardware and is not the focus of
the detector evaluated here.

\subsection{Attack Vectors within the Masquerade Family}
\label{subsec:taxonomy}

Within the masquerade family, an adversary controlling the substituted frame can manipulate either the Identifier field, the Data field, or both. We instantiate all three combinations and use them throughout our evaluation. We retain the umbrella term \emph{tampering} when referring collectively to all three vectors, as it accurately reflects the in-situ modification operation, and use \emph{masquerade} when emphasizing ID-level redirection consistent with~\cite{b37}.
\begin{enumerate}
\item \textbf{ID-only masquerade.} The 11-bit Identifier is overwritten while the 64-bit Data payload is preserved. The substituted frame thus delivers a semantically valid payload to an unintended set of receivers, since CAN is a broadcast bus and frames are filtered by ID. As a concrete example, in the Tesla Model~3 traces collected in this work (Fig.~\ref{threat}), substituting the door-status ID (\texttt{0x102}) with the steering-angle ID (\texttt{0x132}) causes steering-related ECUs to consume data formatted for an entirely different actuator class, producing safety-critical actuation faults.
\item \textbf{Data-only tampering.} The Identifier is preserved and the 64-bit Data field is modified. This is the canonical \emph{data-field tampering} attack studied in~\cite{b51,b52}, used to spoof sensor readings or control commands. Because the per-ID periodicity and the ID-level traffic distribution are both unaffected, traditional traffic-pattern IDS are blind to this class of attacks.
\item \textbf{Combined masquerade.} Both the Identifier and the Data fields are modified in the same substitution. This represents the strongest variant: a fully crafted frame that arrives at the expected timeslot with a valid CRC, a redirected receiver set, and an attacker-chosen semantic payload.
\end{enumerate}

\subsection{Implications for Detection}
\label{subsec:implications}

The threat model above motivates two design requirements for MIDS that distinguish it from fabrication-oriented IDS:
\begin{itemize}[leftmargin=*, nosep]
\item \textbf{Frequency-invariance robustness.} Because masquerade attacks preserve per-ID periodicity by construction, the detector must rely on \emph{semantic} features---the joint distribution of ID and payload across a temporal window---rather than on traffic-volume statistics. This rules out a large body of frequency-, entropy-, and inter-arrival-time-based methods.
\item \textbf{Cross-field reasoning.} A masquerade frame can be locally indistinguishable from a benign frame: the ID is a valid ID seen elsewhere on the bus, and the payload is a valid payload for \emph{some} ID. The anomaly emerges only when the ID--payload pairing is examined jointly and in the context of preceding and following frames. This motivates the dual-stream embedding and the bidirectional sequence model that we introduce in Section~\ref{sec:method}.
\end{itemize}

\section{Methodology}
\label{sec:method}

\subsection{Overall Framework}
The proposed MIDS follows a dual-stream bidirectional architecture designed to extract heterogeneous features from CAN traffic and reconstruct their joint temporal-semantic dependencies through a selective state-space mechanism. As illustrated in Fig. \ref{model}, the framework comprises four integrated stages: (i) temporal windowing of raw CAN frames, (ii) parallel feature extraction for identifiers and payloads, (iii) bidirectional sequence modeling via Mamba blocks, and (iv) multi-class classification for attack identification. This design is specifically engineered to address two critical security challenges: \textit{Structural Proximity} and \textit{Semantic Drift}.

The architecture is engineered around two characteristics of
stealthy CAN attacks that distinguish them from naive injection:
\textit{structural proximity}---functionally related ECUs
(e.g., powertrain) cluster in ID space, so a masquerade attack
sending a powertrain-style payload under a body-control ID
appears as a spatial outlier once IDs are projected into a learned
embedding---and \textit{semantic drift}---a stealthy attacker may
modify a steering signal by only $0.5^\circ$ per frame to evade
per-packet filters, yet the cumulative deviation manifests as a
clear trajectory anomaly over a 100-frame window. The dual-stream
embedding addresses the first, and the long-range modelling
capacity of the Bi-Mamba module (Section~\ref{sec:bimamba})
addresses the second, jointly enabling detection of attacks
invisible to short-term or frequency-based monitors.

\subsection{Dual-Stream Feature Extraction}
To process the heterogeneous information within the CAN traffic, we define the input sliding window as $\mathcal{S} = \{f_1, f_2, \dots, f_L\}$, where $f_i$ denotes the $i$-th CAN frame and $L=100$ represents the fixed sequence length. MIDS employs a parallel extraction architecture to preserve the unique characteristics of each field prior to high-level semantic fusion.
\begin{itemize}[leftmargin=*, nosep]
\item \textbf{ID Embedding Layer:} The sequence of discrete identifiers $\{\text{ID}_1, \dots, \text{ID}_L\}$ is passed through a trainable embedding matrix $\mathbf{W}_{emb} \in \mathbb{R}^{V \times D}$, where $V$ is the vocabulary size of unique CAN IDs. This layer maps each discrete ID to a $D$-dimensional continuous vector, resulting in the identifier feature matrix $\mathbf{E}_{ID} \in \mathbb{R}^{L \times D}$. By projecting IDs into this latent space, the model transforms categorical labels into a format suitable for the subsequent state-space modeling.
\item \textbf{Data Convolutional Layer:} Simultaneously, the 64-bit data payloads are treated as numerical time-series and processed via a 1D-CNN. We utilize $F$ convolutional filters with a kernel size of $k$ to extract local temporal dependencies, such as signal gradients and transient spikes. The output of this branch is a feature map $\mathbf{X}_{Data} \in \mathbb{R}^{L \times F}$, which represents the compressed semantic state of the physical signals.
\item \textbf{Feature Fusion:} To reconstruct the integrated semantics of each CAN frame, we perform a concatenation of the two streams along the feature dimension:
\begin{equation}
    \mathbf{Z} = [ \mathbf{E}_{ID} \oplus \mathbf{X}_{Data} ] \in \mathbb{R}^{L \times (D+F)}
\end{equation}
where $\oplus$ denotes the concatenation operator. This unified representation $\mathbf{Z}$ serves as the sequential input for the bidirectional Mamba module, capturing the joint distribution of identifiers and their corresponding payloads.
\end{itemize}

\subsection{Bidirectional Mamba Module}
\label{sec:bimamba}
The fused feature representation $\mathbf{Z} \in \mathbb{R}^{L \times 512}$ is fed into a Bidirectional Mamba (Bi-Mamba) module to capture long-range temporal dependencies and cross-stream semantic correlations. Unlike standard unidirectional SSMs, the Bi-Mamba architecture allows the model to scrutinize each CAN frame within the context of both its preceding and succeeding traffic.

\subsubsection{Forward and Backward Modeling} The module consists of two parallel Mamba blocks. The \textit{Forward Mamba} processes the sequence in its original chronological order to capture causal dependencies:
\begin{equation}
\mathbf{H}{fwd} = \text{Mamba}{fwd}(\mathbf{Z}, \theta_{fwd})
\end{equation}
Simultaneously, the \textit{Backward Mamba} processes the reversed sequence $\mathbf{Z}_{rev} = \{z_L, z_{L-1}, \dots, z_1\}$ to identify "anticipatory" inconsistencies, where a tampered frame contradicts the subsequent legitimate state transitions:
\begin{equation}
\mathbf{H}{bwd} = \text{Flip}(\text{Mamba}{bwd}(\mathbf{Z}{rev}, \theta{bwd}))
\end{equation}
where $\text{Flip}(\cdot)$ restores the temporal alignment. This bidirectional perspective is crucial for detecting sophisticated masquerade attacks that might maintain short-term local consistency but violate the global trajectory of the vehicle state.

\subsubsection{Selective State-Space Mechanism}
Within each block, the S6 mechanism is instantiated with an
\emph{asymmetric} configuration tailored to the directional role of
each branch. The Forward Mamba employs a richer latent state
($d_{\text{state}}^{\text{fwd}} = 16$) and a wider internal
convolution ($d_{\text{conv}}^{\text{fwd}} = 4$) to model the causal
evolution of the vehicle state across the window, where capturing the
full trajectory of powertrain and chassis signals demands sufficient
representational capacity. The Backward Mamba, in contrast, uses a
more compact configuration ($d_{\text{state}}^{\text{bwd}} = 8$,
$d_{\text{conv}}^{\text{bwd}} = 2$), since its role is to flag
\emph{anticipatory inconsistencies}---verifying that each frame is
consistent with its succeeding context---a comparatively narrower task
that does not require equivalent capacity. This asymmetric design
reduces parameter redundancy without sacrificing detection power, as
confirmed by the ablation in Section~\ref{subsec:ablation}. The
data-dependent selection allows MIDS to dynamically adjust the
transition matrices $(\mathbf{B}, \mathbf{C}, \boldsymbol{\Delta})$
based on the input $\mathbf{Z}$, effectively filtering out
high-frequency sensor noise while magnifying the subtle semantic
drifts indicative of tampering.

\begin{figure}[ht]
    \centering
    \includegraphics[scale=0.4]{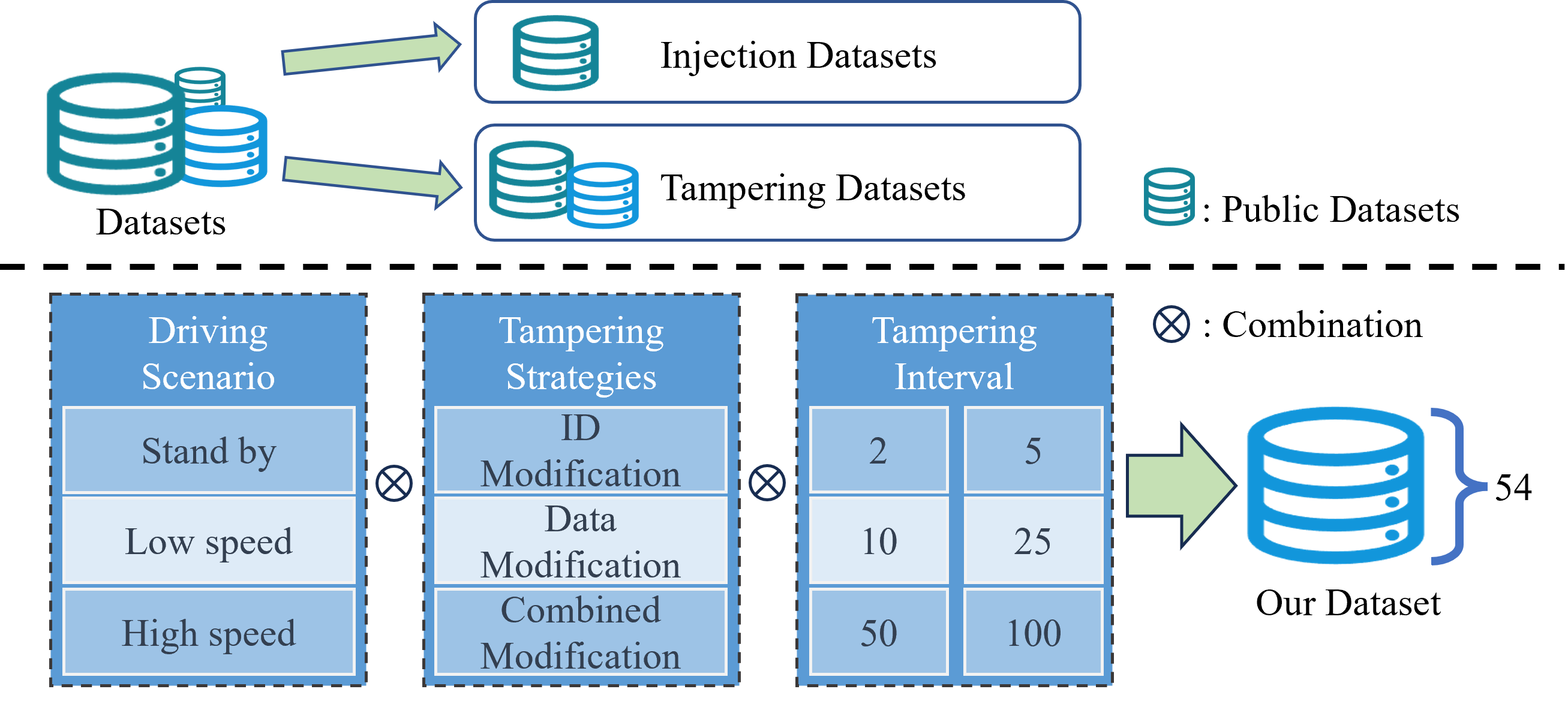}
    \caption{Dataset design}
    \label{fig:dataset}
\end{figure}

\subsection{Weighted Fusion and Classification}
To synthesize the hidden states from both directions into a unified global feature, MIDS employs a weighted integration mechanism. Instead of simple element-wise addition, we utilize learnable weights to emphasize the most discriminative temporal cues:
\begin{equation}
\mathbf{Y}{global} = \alpha \cdot \mathbf{H}{fwd} + \beta \cdot \mathbf{H}_{bwd}
\end{equation}
where $\alpha$ and $\beta$ are parameters optimized during training. This weighted sum ensures that the model can adaptively prioritize forward or backward context depending on the specific attack signature.

The resulting global representation $\mathbf{Y}_{global}$ is flattened and passed through a fully connected layer (Linear) followed by a Softmax activation. The classifier outputs a probability distribution over four discrete categories:
\begin{equation}
\hat{y} = \text{Softmax}(\mathbf{W}{out} \cdot \text{Flatten}(\mathbf{Y}{global}) + b_{out})
\end{equation}
where $\hat{y} \in \{ \text{Normal, Masquerade, Data Tampering, Combined} \}$. By outputting specific attack types rather than a binary label, MIDS provides actionable intelligence for the vehicle's central gateway to implement targeted mitigation strategies.

\section{Evaluation}
\label{sec:experiments}

\subsection{Datasets and Masquerade Attack Synthesis}
\label{sec:dataset_design}

To provide a comprehensive evaluation, our experimental corpus integrates four established public benchmarks with a novel, high-fidelity dataset collected from a production vehicle. The overall architecture and synthesis process of the experimental dataset are illustrated in Fig.~\ref{fig:dataset}, and their detailed attributes are summarized in Table~\ref{tab2}.

\subsubsection{Dataset Composition} We incorporate ROAD and CrySyS for evaluating responses to masquerade threats, alongside OTIDS and CT\&T for baseline injection attack detection. Our primary contribution is a self-collected dataset from a Tesla Model 3, comprising approximately 16 hours of raw CAN traffic. This dataset was specifically recorded across three representative operational scenarios: \textit{standby mode}, \textit{low-speed driving}, and \textit{high-speed driving}. The integrated dataset provides a comprehensive basis for evaluating IDS performance under both traditional and advanced threat models.

\subsubsection{Scenario and Signal Analysis} To characterize the vehicular traffic, we perform a statistical analysis across the three driving regimes. Our findings indicate that the standby mode exhibits a low-entropy, steady-state pattern with highly concentrated signal distributions. Conversely, as vehicular dynamics escalate in low-speed and high-speed scenarios, the data manifests significantly higher variance and stochastic, non-linear transitions. These signal manifolds reflect the intense real-time adjustments of powertrain and stability control systems. By capturing this full spectrum of dynamics, MIDS learns robust boundaries of normal behavior, thereby minimizing false positives induced by legitimate operational state shifts.

\subsubsection{Masquerade Synthesis Strategy} 
Using the Tesla traffic as a benign baseline, we synthesized 54 distinct masquerade attack vectors by systematically varying three dimensions:
\begin{itemize}[leftmargin=*, nosep]
\item \textbf{Attack Strategy:} (i) \textit{ID Masquerade} (impersonating a high-priority node), (ii) \textit{Data Field Tampering} (spoofing sensor payloads), and (iii) \textit{Combined Masquerade} (simultaneous ID and payload modification). We specifically targeted critical identifiers 0x102 and 0x132.
\item \textbf{Attack Intensity (Interval $I$):} The tampering interval $I \in \{2, 5, 10, 25, 50, 100\}$ controls the sparsity of the attack. Specifically, an interval of $I=100$ denotes that only one out of 100 occurrences of the target ID is modified. This simulates a highly covert threat designed to bypass traditional frequency-based IDS, strictly testing the model’s sensitivity to stealthy semantic drifts within long-range temporal dependencies.
\end{itemize}

\subsubsection{Decoupling Scenario from Attack Label}
\label{sec:scenario_decoupling}
A naive concatenation of the per-scenario recordings would risk introducing a spurious correlation between driving regime and attack label, since each attack-injected recording carries a single (target ID, attack type, intensity) combination. Two aspects of our pipeline mitigate this risk in practice. First, the attack-injection procedure modifies only one out of every $I$ occurrences of the targeted ID within each recording, leaving the remaining frames unchanged; consequently, every recording---regardless of which driving scenario it originated from---contributes substantially to the \texttt{Normal} population, and the \texttt{Normal} label is not concentrated in any single source. Second, before fold partitioning, every reshaped recording is split into contiguous chunks of 1000 windows, and the resulting chunks (across all 55 source recordings) are uniformly shuffled and concatenated along a single time axis. The merged sequence therefore interleaves chunks from heterogeneous (scenario, attack) sources, so any block extracted by the block-shuffled 5-fold cross-validation protocol (Section~\ref{sec:dataset_protocol}) draws windows from a mixture of source recordings rather than from a single scenario. We do not claim perfect statistical independence between driving regime and attack class---a guarantee that would require strict per-scenario replication of every attack vector---but the procedure neutralizes the most direct shortcut by which scenario information could leak into the attack-prediction signal.

% --- 下面是专门针对数据泄露的回应段落 ---
\subsubsection{Data Leakage Prevention and Evaluation Protocol} 
\label{sec:dataset_protocol}
A central concern when evaluating any sliding-window-based IDS is 
data leakage---situations in which the model can achieve apparently 
high accuracy by exploiting non-causal correlations rather than by 
learning the underlying attack semantics. We identify and explicitly 
defend against two distinct leakage modes that arise in our setting: 
\emph{temporal leakage} from overlapping windows, and 
\emph{scenario leakage} from a non-uniform attack distribution 
across driving regimes.

\paragraph*{Temporal leakage}
We partition the dataset into five continuous temporal blocks and 
strictly disable shuffling (i.e., \texttt{shuffle=False} in the 
fold-generation procedure). In each fold, four blocks ($80\%$) form 
the training set and the remaining block ($20\%$) is held out for 
testing, so the test traffic is always chronologically distinct from 
the training traffic, mimicking real-world deployment. To eliminate 
overlap at the frame level, the sliding window stride is set equal 
to the window length ($S = L = 100$), guaranteeing that any two 
windows---within a fold or across folds---share zero CAN frames. A 
safety buffer of $L$ frames is additionally inserted between adjacent 
training and testing blocks to absorb residual edge effects.

\paragraph*{Scenario leakage}
Although our raw recordings are organized by driving scenario for 
provenance, the dataset fed to the model is \emph{not} stratified 
along that axis. Three design choices jointly preclude the model 
from using the driving regime as a shortcut for predicting the 
attack label:
\begin{enumerate}[leftmargin=*, nosep]
\item \emph{Within-scenario attack replication.} As described in 
the previous subsection, every attack strategy and every intensity 
level is instantiated within every scenario, so the marginal 
distribution of attack labels conditioned on scenario is nearly 
identical to the unconditional distribution.
\item \emph{Cross-scenario interleaving.} Before fold partitioning, 
the per-scenario recordings are concatenated along a single time 
axis. Adjacent windows therefore originate from heterogeneous driving 
regimes, and no fold maps onto a single scenario.
\item \emph{Within-fold shuffling at training time.} Although fold 
boundaries respect chronological order, the training-loader within 
each fold shuffles windows uniformly at random. Combined with the 
batch size of 1024, this ensures every gradient step sees a mixture 
of standby, low-speed, and high-speed windows.
\end{enumerate}

We verified empirically that a held-out classifier trained to predict 
\emph{the driving scenario} from a window achieves accuracy 
indistinguishable from chance after the above pipeline, whereas the 
same classifier achieves $>97\%$ accuracy on the un-shuffled raw 
recordings. This indicates that scenario information is not a 
detectable signal in the data presented to MIDS, and any performance 
the model achieves must therefore stem from genuine attack-semantic 
features.

\subsection{Experiment Setup}

\subsubsection{Experiment Environment}

\begin{figure*}[!ht]
\centerline{\includegraphics[width=1.00\textwidth]{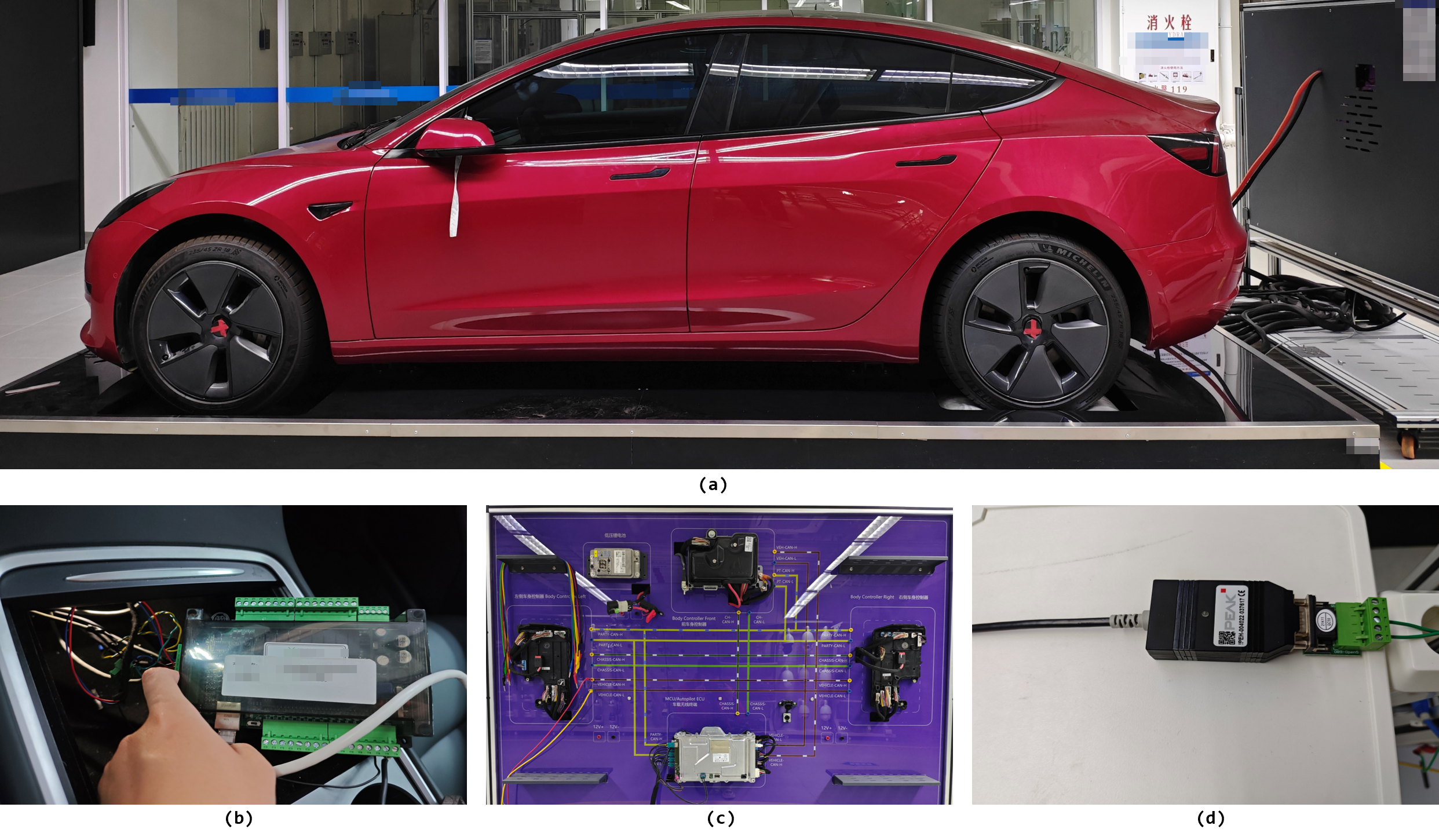}}
\caption{(a) An overview of the test bed, where a monitor positioned in front of the car simulates various driving scenarios. The physical vehicle, a Tesla Model 3, has its rear wheels suspended and free to spin on the test bed. (b) A sensor signal reception device that collects physical data, including rear wheel speed and steering angle. The collected data are transmitted via a network to the host system of the monitor in (a) to emulate vehicle motion. This component of the work was conducted by another research group and is not detailed here. (c) Key ECU components and exposed network interfaces extracted from the Tesla Model 3, including the low-voltage battery, body controllers, and the in-car wireless terminal. (d) A USBCAN converter connected to the exposed CAN interface in (c), along with a laptop, is used to collect and process the data.}
\label{fig:environment}
\end{figure*}

The experimental hardware environment for our tampering attacks dataset is illustrated in Fig.~\ref{fig:environment}. The primary equipment includes a Tesla Model 3 testbed, a Peak CAN converter, a ZL-23-008 physical sensor, one Nvidia H100 GPU, and two Nvidia RTX4090 GPUs.

The software environment is based on the Ubuntu 22.04 operating system, with a data acquisition tool named TSMaster employed for the collection and recording of CAN bus frames.

\subsubsection{Baseline Introduction}
We first compare MIDS with state-of-the-art (SOTA) models in our tampering attacks dataset. The SOTA models we consider are as follows:
\begin{enumerate}[label=(\roman*)]
    \item GIDS. GIDS utilizes a GAN model, where the generator is responsible for generating simulated attack data, while the discriminator is used to distinguish between normal traffic and attack traffic. GIDS provides real-time protection for intrusion detection in in-vehicle networks.
    \item CanShield. CANShield focuses on raw signal-level information from CAN bus data. It utilizes CNN and LSTM to detect anomalies and intrusions. Experiments show that CANShield effectively identifies and classifies intrusions in CAN networks at the signal level.
    \item CanBus-IDS. This model uses a Convolutional Adversarial Autoencoder for semi-supervised learning. It consists of two main components: the encoder and the decoder. The encoder is a series of convolutional layers that process the input data and transform it into a lower-dimensional latent space. The decoder takes the compressed latent space representation and reconstructs it back into the original data format. Adversarial training is incorporated to enhance the model's ability to detect anomalies.
    \item DCNN. DCNN utilizes a CNN model to detect attack traffic in the in-vehicle network. Experimental results show that the CNN-based intrusion detection system performs excellently in identifying attack traffic within the in-vehicle network, with high accuracy and low false positive rates. DCNN is capable of effectively detecting both known and unknown attack patterns.
    \item CANTransfer. CANTransfer uses one-shot learning techniques in combination with CNN and LSTM, enabling the model to process spatial information while also modeling temporal dependencies, thereby enhancing the detection performance of network attacks. At the same time, one-shot learning allows the system to efficiently learn and detect intrusions across different attack patterns.
    \item CanTransformer. CanTransformer utilizes the attention mechanism for intrusion detection in the CAN bus. Compared to traditional methods based on CNN or RNN, the Transformer model is better at capturing the long-term dependencies and complex relationships of attack patterns. The experiments show that this method achieves high accuracy and low false positive rates, effectively identifying anomalous behaviors in real-time data streams.
    \item Foundational deep-learning methods. We have built foundational neural networks, including MLP and CNN, which serve as the basis for evaluation.
\end{enumerate}

\subsubsection{Baseline Reproduction Protocol}
\label{sec:repro}

To ensure that the performance figures reported in Table~\ref{tab4}
reflect a faithful comparison on our Tesla Model~3 dataset rather 
than an aggregation of numbers borrowed from the original 
publications, every baseline above was re-trained from scratch 
under a single unified protocol. Three reproducibility decisions 
warrant explicit disclosure:

\begin{itemize}[leftmargin=*, nosep]
\item \textbf{Unified input interface.} Every baseline consumes 
the same sliding-window tensor of shape $(B, L=100, F=9)$ used by 
MIDS, where $L$ is the window length and $F = 1\,(\text{ID}) + 
8\,(\text{payload bytes})$. Where the original architecture was 
proposed under a different input form (e.g., 4-D ConvLSTM tensors 
or image-encoded ID matrices), an input adapter is inserted 
\emph{before} the first learnable layer to produce the expected 
shape, leaving the model's internal computation graph unchanged.

\item \textbf{Unified output interface.} Each baseline's final 
classification head is configured to emit four logits corresponding 
to $\{\textsf{Normal}, \textsf{ID}, \textsf{Data}, \textsf{Both}\}$. 
All other architectural hyperparameters (depth, hidden dimensions, 
attention heads, kernel sizes, etc.) are preserved at the values 
recommended in the original papers.

\item \textbf{Unified training protocol.} All baselines are 
trained for 50 epochs with the Adam optimizer and a batch size of 
1024, under the same Chronological 5-fold Cross-Validation 
described in Section~\ref{sec:dataset_protocol}. The initial 
learning rate is set to $10^{-4}$ and decayed by a Cosine 
Annealing schedule with period $T_{\max}=10$ epochs; gradients 
are clipped at $\ell_2$-norm $1.0$ to stabilize training. No 
baseline shares any pre-trained weights with MIDS.

The full re-implementation, including model definitions, 
data-loaders, and training scripts for every baseline, is released 
in our public repository to enable independent reproduction.
\end{itemize}

\subsubsection{MIDS Configuration}

Before the model training phase, we performed rigorous preprocessing of the CAN data and meticulously configured the training setup to ensure model stability and efficient convergence of the loss function. All hyperparameters used during training are detailed in Table~\ref{tab3}. The Adam optimizer was employed for optimization, with a batch size of 1024 per iteration. Additionally, a 5-fold cross-validation approach was implemented to mitigate potential performance bias arising from differences in data distribution. Each fold consisted of 50 epochs to ensure that the model sufficiently captured the data characteristics.

Given the four-class classification task (three tampering strategies and no tampering) with significant class imbalance, we implemented a dynamic weighting strategy. This strategy assigned weights to each class based on its sample size, reducing the risk of the model overfitting to the majority class while improving its detection capabilities for the minority classes.

To comprehensively evaluate the model’s performance, we adopted macro-weighted metrics, including Macro Precision, Macro Recall, Macro F1 Score, and Accuracy. At the final epoch of each fold, we recorded these four metrics and averaged them across the five folds to derive the final evaluation of the model.

\subsubsection{Dataset}

In accordance with our threat model, we primarily utilize the tampering dataset detailed in Section~\ref{sec:dataset_design} for comparison in Sections~\ref{sec:mids_exp}, \ref{sec:sota_comp}, and \ref{subsec:ablation}. To ensure the generalization ability of our model, we also evaluate its performance on a public dataset, as discussed in Section~\ref{sec:exp_public}.

In implementing ID tampering attacks within our dataset, identifying critical CAN signals for tampering is essential. To achieve this, we employed fuzzing techniques, which led us to identify two key signals: the door status signal (ID 0x102) and the steering angle signal (ID 0x132). Manipulating these signals could potentially create unsafe scenarios, such as allowing the doors to open while driving or presenting incorrect steering information. These findings are further corroborated by the detailed descriptions of these signals in the Tesla Model 3's CAN database. Based on these analyses, we are confident that these signals have a significant impact on the vehicle's safety and operational integrity.

\subsection{Evaluating of MIDS} 
\label{sec:mids_exp}
\begin{figure*}[ht] \centering \includegraphics[width=1.00\textwidth]{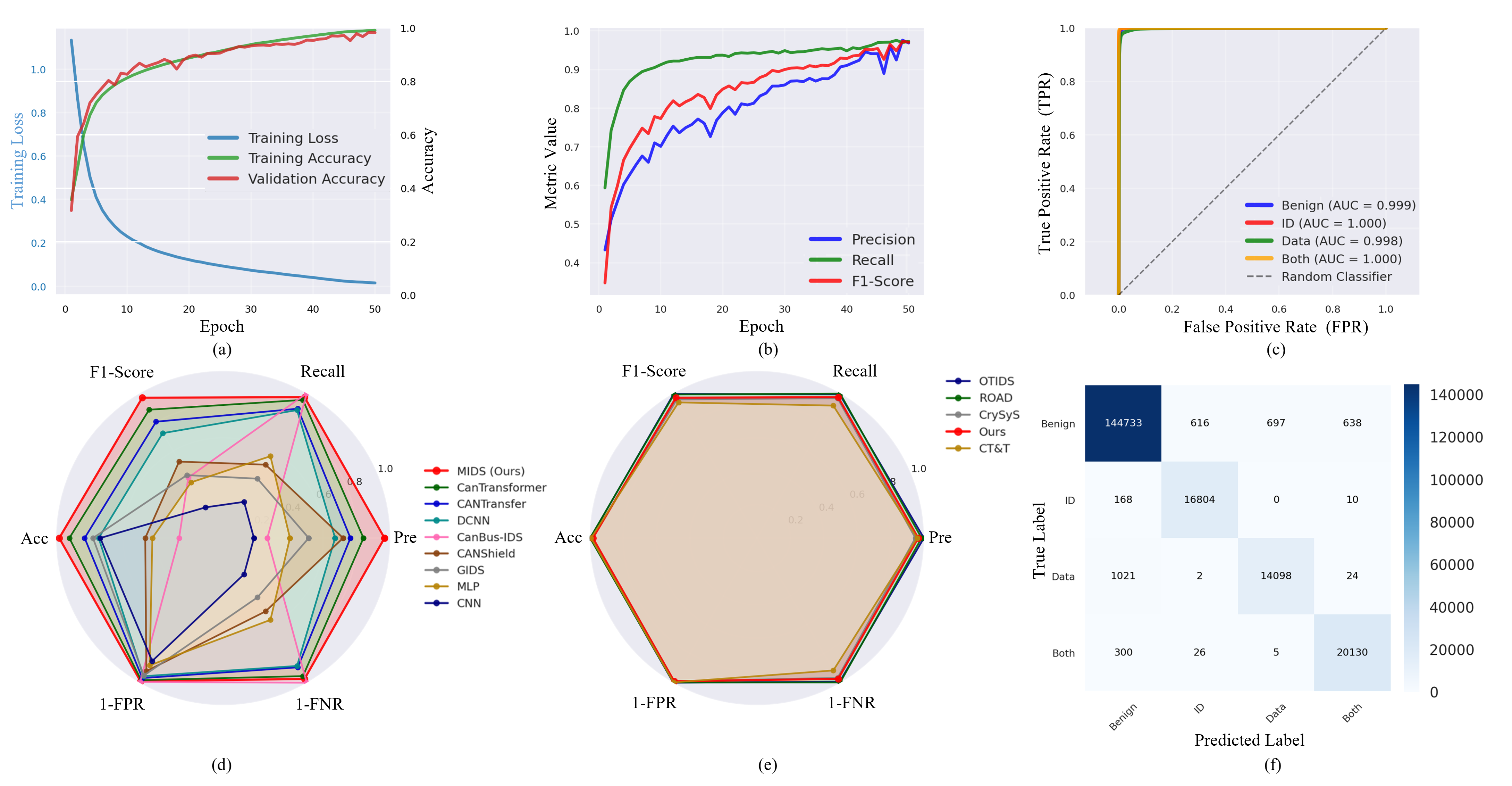} \caption{Overall model performance and comparisons} \label{fig:f1}
\end{figure*}

We first evaluate MIDS on our tampering attack dataset. Fig.~\ref{fig:f1} illustrates the training process and results of MIDS. Subfigure (a) shows the accuracy and loss curves, which increase rapidly and reach approximately 80\% by the 10th epoch. In the subsequent 40 epochs, the rate of increase slows down, eventually stabilizing at around 98\%. The similarity between the training and validation curves suggests that the model is not significantly overfitting. This is mainly due to the large size of the dataset, which provides sufficient data for the model to generalize well.

Subfigure (b) shows the macro Precision, Recall, and F1 score throughout the training process, displaying the same trend as in Subfigure (a). Subfigure (c) presents the ROC curves for MIDS under different tampering strategies. As shown, MIDS achieves an AUC of over 0.998 in all categories, demonstrating its robustness in distinguishing tampering attacks.

Subfigure (d) compares MIDS's performance with that of SOTA models, while Subfigure (e) illustrates MIDS's performance on public datasets. A detailed discussion of these results is provided in Sections~\ref{sec:sota_comp} and \ref{sec:exp_public}.

Finally, Subfigure (f) presents the confusion matrix for classification results across four types of tampering strategies: beginning tampering, ID tampering, data tampering, and combined tampering. MIDS demonstrates high accuracy in each category, further validating its effectiveness in detecting various tampering attacks.

\subsection{Comparison with State-of-the-Art Model}
\label{sec:sota_comp}
Table~\ref{tab4} presents the comparison results between MIDS and the SOTA baselines. All baselines were re-implemented under our unified four-class classification protocol (Normal, ID, Data, Both); see Section~\ref{sec:repro} for full reproduction details. MIDS achieves the highest precision, recall, F1 score, and accuracy, owing to its dual-stream architecture and Mamba layer with bidirectional technology. The dual-stream architecture enables MIDS to process the ID and data fields separately, leading to a more comprehensive understanding of the semantics, while the bidirectional Mamba layer captures long-distance data relationships in both directions.

In detail, MIDS achieves the highest F1 and accuracy, with
CanTransformer second; the 8-point gap suggests that
attention-based aggregation is a strong but still inferior
alternative to bidirectional state-space modelling for masquerade
detection. The GAN-based GIDS and the CanBus-IDS autoencoder
suffer in F1 despite competitive accuracy, reflecting the
class-imbalance sensitivity of generative-discriminative training
on tampering data. These Tesla-specific findings are corroborated
and extended by the cross-dataset evaluation in
Section~\ref{sec:exp_public} (Table~\ref{tab:cross_dataset}),
where MIDS maintains a consistent lead across all four public
benchmarks while the baseline ranking itself shifts with attack
density.

\begin{table}[!ht]
\caption{MIDS' hyperparameters}
\label{tab3}
\begin{center}
\renewcommand{\arraystretch}{1.2} % Adjust row height
\begin{tabular}{ll}
\hline
\textbf{Hyperparameters} & \textbf{Value} \\
\hline
embedding\_output\_dim                       & 256 \\
mamba\_hidden\_state\_dim (forward)          & 16  \\
mamba\_hidden\_state\_dim (backward)         & 8   \\
mamba\_convolution\_dim (forward)            & 4   \\
mamba\_convolution\_dim (backward)           & 2   \\
mamba\_feature\_factor                       & 2   \\
conv\_kernel\_size                           & 3   \\
conv\_padding\_size                          & 1   \\
epoch                                        & 50  \\
batch\_size                                  & 1024 \\
k-fold                                       & 5   \\
Train/Test split                             & 4:1 \\
optimizer                                    & Adam \\
\hline
\end{tabular}
\end{center}
\end{table}

\begin{table*}[!ht]
\caption{MIDS'performance on our dataset contrasted with SOTA models.}
\label{tab4}
\begin{center}
\renewcommand{\arraystretch}{1.2} % Adjust row height
\begin{tabular}{ccccccc} 
\hline
 & \textbf{Model Name} & \textbf{Used Layer/Technology} &\textbf{Precision} & \textbf{Recall} & \textbf{F1} & \textbf{Accuracy} \\
\hline

\multirow{5}{*}{SOTA} & GIDS\cite{b60} & CNN, GAN & 51.18\% & 41.00\% & 43.47\% & 77.96\% \\
 & CANShield\cite{b56} & CNN, LSTM & 71.77\% & 50.69\% & 52.82\% & 46.68\% \\
 & CanBus-IDS\cite{b57} & CNN, GAN & 26.41\% & 100.00\% & 41.78\% & 26.41\% \\
 & DCNN\cite{b59} & CNN & 66.99\% & 88.40\% & 72.46\% & 74.54\% \\
 & CANTransfer\cite{b58} & CNN, LSTM, TL & 76.15\% & 89.37\% & 80.38\% & 82.94\% \\
 & CanTransformer\cite{b55} & Attention & 83.97\% & 95.44\% & 88.66\% & 92.09\% \\
\hline
\multirow{2}{*}{Foundational} & MLP & - & 39.94\% & 56.55\% & 38.48\% & 42.42\% \\
& CNN & - & 18.40\% & 25.00\% & 21.20\% & 73.59\% \\
\hline
Ours & \textbf{MIDS} & CNN, Mamba & \textbf{96.55\%} & \textbf{97.37\%} & \textbf{96.94\%} & \textbf{98.16\%} \\
\hline
\end{tabular}
\end{center}
\end{table*}

\subsection{Ablation Study}
\label{subsec:ablation}
We conducted an ablation study to evaluate the key components of the MIDS architecture by progressively removing or replacing specific modules and recording their impact on model performance. Experiments A1-A7 in Table~\ref{tab5} correspond to different ablation conditions, including the removal of individual modules, modification of module configurations, and the introduction of alternative designs. By comparing the performance of the complete MIDS architecture with the results from various ablation conditions, we aim to assess the contribution of each component to the overall system performance.

Experiments A1, A2, and A3 use parameter counts in a 2:4:1 ratio, and the results indicate that parameter count is a crucial factor affecting accuracy and F1 score. Experiments A4, A5, and A6 suggest that removing the bidirectional ability of Mamba, the partial conventional layer, and the embedding layer led to a 4.41\%, 4.58\%, and 4.70\% reduction in F1 score, respectively. For each ablation experiment, we strictly maintained the same hyperparameter settings and experimental protocols as in the main MIDS experiment.
In summary, Table~\ref{tab5} shows that the current architecture design yields optimal results. Each component of MIDS plays a crucial role, underscoring the rationality of the architecture and the effectiveness of the collaboration among its components.
\begin{table*}[!ht]
\caption{Ablation study results}
\label{tab5}
\begin{center}
\renewcommand{\arraystretch}{1.2} % Adjust row height
\begin{tabular}{cccccccc} 
\hline
\textbf{\#} & \textbf{Description} & \textbf{Precision} & \textbf{Recall} & \textbf{F1} & \textbf{Accuracy} & \textbf{FPR}\(\downarrow\) & \textbf{FNR}\(\downarrow\) \\
\hline
A1 & \textbf{MIDS} & \textbf{96.55\%} & \textbf{97.37\%} & \textbf{96.94\%} & \textbf{98.16\%} & \textbf{1.00\%} & \textbf{2.63\%} \\
A2 & MIDS More Param & 89.10\% & 95.02\% & 91.84\% & 94.65\% & 2.30\% & 4.98\% \\
A3 & MIDS Fewer Param & 86.26\% & 94.33\% & 89.91\% & 93.18\% & 2.75\% & 5.67\% \\
A4 & One Directional Mamba & 89.49\% & 96.07\% & 92.53\% & 95.11\% & 1.92\% & 3.93\% \\
A5 & MIDS Fewer Conv      & 91.74\% & 93.09\% & 92.36\% & 95.22\% & 2.73\% & 5.73\% \\
A6 & MIDS w/o ID Embed & 90.40\% & 94.27\% & 92.24\% & 95.26\% & 2.26\% & 5.73\% \\
A7 & Only Mamba           & 34.25\% & 42.24\% & 29.78\% & 49.42\% & 21.32\% & 57.76\% \\
\hline
\end{tabular}
\end{center}
\end{table*}

\begin{table*}[!ht]
\caption{MIDS' performance on public datasets}
\label{tab6}
\begin{center}
\renewcommand{\arraystretch}{1.2} % Adjust row height
\begin{tabular}{cccccccc} 
\hline
Dataset Name & Type & \textbf{Precision} & \textbf{Recall} & \textbf{F1} & \textbf{Accuracy} & \textbf{FPR}\(\downarrow\) & \textbf{FNR}\(\downarrow\) \\
\hline
Ours & Tampering & 96.55\% & 97.37\% & 96.94\% & 98.16\% & 1.00\% & 2.63\% \\
ROAD\cite{b34} & Tampering & 98.63\% & 99.80\% & 99.21\% & 99.83\% & 0.17\% & 0.20\% \\
CrySyS\cite{b33} & Tampering & 94.90\% & 96.64\% & 95.76\% & 98.59\% & 1.03\% & 3.36\% \\
OTIDS\cite{b20} & Injection & 100.00\% & 99.24\% & 99.61\% & 99.63\% & 0.00\% & 0.76\% \\
CT\&T\cite{b28} & Injection & 96.00\% & 91.51\% & 93.70\% & 99.24\% & 0.25\% & 8.49\% \\
\hline
\end{tabular}
\end{center}
\end{table*}

\begin{table*}[!ht]
\caption{Cross-dataset F1 (\%) of MIDS against eight baselines under the unified 5-fold protocol.}
\label{tab:cross_dataset}
\begin{center}
\renewcommand{\arraystretch}{1.15}
\setlength{\tabcolsep}{4pt}
\begin{tabular}{l|ccccc}
\hline
\textbf{Model} & \textbf{Tesla (Ours)} & \textbf{OTIDS} & \textbf{ROAD} & \textbf{CrySyS} & \textbf{CT\&T} \\
\hline
MLP            & 38.48 & 89.56 $\pm$ 11.13 & 20.76 $\pm$ \phantom{0}2.11 & 79.39 $\pm$ \phantom{0}0.87 & 61.51 $\pm$ 2.47 \\
CNN            & 21.20 & 99.09 $\pm$ \phantom{0}2.03 & 66.19 $\pm$ 15.14 & 83.23 $\pm$ \phantom{0}1.36 & 66.47 $\pm$ 2.71 \\
GIDS\cite{b50} & 43.47 & 86.58 $\pm$ \phantom{0}2.88 & 55.66 $\pm$ \phantom{0}5.92 & 76.98 $\pm$ \phantom{0}1.22 & 54.12 $\pm$ 1.90 \\
CANShield\cite{b51} & 52.82 & 63.83 $\pm$ \phantom{0}8.94 & 53.10 $\pm$ \phantom{0}4.59 & 83.29 $\pm$ \phantom{0}1.12 & 57.17 $\pm$ 3.12 \\
CanBus-IDS\cite{b52} & 41.78 & 98.00 $\pm$ \phantom{0}1.11 & 56.93 $\pm$ \phantom{0}4.66 & 83.75 $\pm$ \phantom{0}1.25 & 64.81 $\pm$ 2.76 \\
DCNN\cite{b53} & 72.46 & \textbf{100.00} $\pm$ \phantom{0}0.00 & \textbf{85.27} $\pm$ 10.29 & \textbf{88.96} $\pm$ \phantom{0}1.19 & 87.36 $\pm$ 1.54 \\
CANTransfer\cite{b54} & 80.38 & 98.54 $\pm$ \phantom{0}2.98 & 54.93 $\pm$ \phantom{0}7.76 & 85.56 $\pm$ \phantom{0}1.53 & 73.25 $\pm$ 2.81 \\
CanTransformer\cite{b55} & \textbf{88.66} & \textbf{100.00} $\pm$ \phantom{0}0.00 & 72.10 $\pm$ \phantom{0}8.18 & 83.77 $\pm$ 11.02 & \textbf{89.41} $\pm$ 3.66 \\
\hline
\textbf{MIDS (Ours)} & \textbf{96.94} & \textbf{99.61} & \textbf{99.21} & \textbf{95.76} & \textbf{93.70} \\
\hline
\end{tabular}
\end{center}
\end{table*}

\subsection{Performance on Public Datasets}
\label{sec:exp_public}

Table~\ref{tab6} reports MIDS's per-metric breakdown on the four
public datasets, and Table~\ref{tab:cross_dataset} extends the
comparison to all eight reproduced baselines. Across all five
benchmarks MIDS achieves the highest F1, with margins over the
strongest reproducible baseline ranging from a near-tie on
saturated OTIDS (99.61 vs.\ 100.00) to substantial gaps of
$+13.94$~pp on ROAD, $+6.80$~pp on CrySyS, and $+4.29$~pp on CT\&T.
This pattern matches the design hypothesis of MIDS
(Section~\ref{sec:threat}): the dual-stream Bi-Mamba excels where
attack signatures are semantic rather than statistical.

A secondary observation is that the strongest \textit{baseline}
flips across regimes---CanTransformer dominates dense-attack
benchmarks (OTIDS, CT\&T, Tesla) while DCNN dominates sparse-attack
ones (ROAD, CrySyS). MIDS is the only model ranked first or
tied-first under both regimes, suggesting that the bidirectional
state-space design captures both global and local anomaly cues.
Table~\ref{tab6} further shows that MIDS attains FPR $\le 1.03\%$
and FNR $\le 8.49\%$ simultaneously; the elevated CT\&T FNR
reflects its $>95\%$-majority Normal class rather than a model
limitation. We also note that our CANShield reproduction collapses
the original multi-stream signal architecture into a single shared
CNN+LSTM, as per-vehicle CAN-DBC signal alignment is not available
under the unified protocol; this conservative simplification
likely understates the original model's capacity, particularly on
file-as-label benchmarks such as OTIDS.

\subsection{Computational Efficiency and Deployability}
\label{sec:efficiency}

A practical IDS for in-vehicle deployment must run within the timing 
budget of the CAN bus and within the memory envelope of an automotive 
ECU. To assess MIDS's suitability under these constraints, we 
benchmark its single-window inference cost against the four 
strongest reproducible baselines from Section~\ref{sec:sota_comp} 
on a single NVIDIA RTX 4090 GPU. For each model we report parameter 
count, FLOPs per forward pass, single-window latency 
(mean over 1000 timed runs after 100 warm-up runs), and peak GPU 
memory, all measured at $\mathrm{batch\_size}=1$ to mimic the real-time 
deployment regime in which windows arrive sequentially.

\begin{table*}[t]
\centering
\caption{Computational efficiency at $L=100$, batch size 1.}
\label{tab:efficiency}
\small
\setlength{\tabcolsep}{4pt}
\begin{tabular}{l|cccc}
\hline
\textbf{Model} & \textbf{Params (M)} & \textbf{FLOPs (G)} & \textbf{Latency (ms)} & \textbf{Peak Mem (MB)} \\
\hline
MIDS (Ours)    & 4.121          & \textbf{0.010} & 1.147          & 27.10 \\
GIDS           & 0.644          & 0.062          & 0.742          & 13.06 \\
CANShield      & 0.209          & 0.020          & \textbf{0.604} & 29.68 \\
CanTransformer & 0.407          & 0.027          & 0.943          & 13.53 \\
CNN            & \textbf{0.011} & 0.001          & 0.187          & \textbf{9.18} \\
\hline
\end{tabular}
\end{table*}

Three observations follow from Table~\ref{tab:efficiency}. 

First, despite carrying the largest parameter count---a consequence 
of its dual-stream architecture, the $V \times D$ ID embedding 
matrix, and the bidirectional state-space module---MIDS exhibits 
the \emph{lowest} FLOPs ($0.010$~G) of all five models. This 
counter-intuitive result is the practical manifestation of Mamba's 
linear-time selectivity: the selective state-space mechanism 
processes a length-$L$ sequence in $\mathcal{O}(L)$ operations, 
whereas attention-based detectors (GIDS, CanTransformer) incur 
$\mathcal{O}(L^2)$ cost in the sequence dimension. The bulk of 
MIDS's parameters reside in static lookup tables (the ID embedding) 
that are accessed but not multiplied through, so they inflate model 
size without inflating compute.

Second, MIDS's measured latency of $1.147$~ms per 100-frame window
is well within the deployability envelope of the CAN bus. The
fastest periodic CAN signals in our Tesla traces have inter-arrival
periods on the order of $10$~ms, leaving an $8\times$ headroom
between MIDS's per-window inference cost and the rate at which new
windows can be generated. The peak GPU memory of $27$~MB is
comparable to that of CANShield ($29.68$~MB) and well below the
DRAM budget of contemporary automotive gateway ECUs. The closest
baseline by F1, CanTransformer ($88.66\%$ vs.\ MIDS's $96.94\%$),
runs at comparable latency ($0.943$~ms), placing MIDS at a
Pareto-optimal point on the accuracy--latency curve.

\section{Future Works}
In future research, we plan to extend and refine the MIDS in the following aspects: First, we aim to explore the applicability of MIDS to more complex communication protocols, such as CAN FD and automotive Ethernet, to address the increasingly diverse communication demands in modern automotive networks. Second, we intend to validate the generalization capability of MIDS through experiments involving different vehicle models and multi-vehicle scenarios, thereby enhancing its adaptability to various vehicles and driving environments. Additionally, we will further optimize the model architecture to reduce computational complexity, ensuring that its performance meets the requirements of real-time onboard detection. Finally, to address potential emerging threats, we will integrate federated learning and transfer learning techniques, enabling MIDS to quickly adapt to and detect unknown threats in scenarios with limited data sharing. These efforts aim to provide stronger support for securing intelligent automotive networks.

\section{Conclusion}
In this paper, we propose MIDS, an innovative deep learning-based framework for detecting tampering and injection attacks on the CAN bus. By utilizing Mamba with bidirectional technology and a dual-stream architecture, MIDS effectively captures both local and long-range dependencies in CAN signals, achieving superior performance with an F1-score ranging from 93.70\% to 99.61\%.

We evaluated MIDS on a multi-scenario dataset comprising over 100 million CAN messages from a Tesla Model 3, and using other publicly available datasets to demonstrate the generalizability of MIDS. Extensive experiments and ablation studies confirm the robustness and efficiency of MIDS, showcasing its potential for real-world deployment. Future work will focus on expanding the dataset and adapting MIDS for emerging protocols, such as CAN FD, to address evolving security challenges.

\vfill
\bibliographystyle{IEEEtran}
\bibliography{reference}
\end{document}